\documentclass[11pt,a4paper]{article}
\pdfoutput=1
\usepackage{graphicx}
\usepackage{amsmath,amsthm,latexsym,amssymb,amsfonts,epsfig}
\usepackage{fontenc,dsfont,layout}
\usepackage{hyperref}
\usepackage{psfrag}
\usepackage[cp1250]{inputenc}
\numberwithin{equation}{section}

\DeclareMathAlphabet{\mathpzc}{OT1}{pzc}{m}{it}

\begin{document}

\title{Vector fields and Loop Quantum Cosmology}
\author{Micha{\l} Artymowski\thanks{Michal.Artymowski@fuw.edu.pl} $\;\,$  and $\;$
           Zygmunt Lalak\thanks{Zygmunt.Lalak@fuw.edu.pl}}
\date{\it  Institute of Theoretical Physics, Faculty of Physics, University of Warsaw ul. Ho\.{z}a 69, 00-681 Warszawa, Poland} 
\maketitle
x

\begin{abstract}
In the context of the Loop Quantum Cosmology we have analysed the holonomy correction to the classical evolution of the simplified Bianchi I model in the presence of vector fields. For the Universe dominated by a massive vector field or by a combination of a scalar field and a vector field a smooth transition between Kasner-like and Kasner-unlike solutions for a Bianchi I model has been demonstrated. In this case a lack of initial curvature singularity and a finite maximal energy density appear already at the level of General Relativity, which simulates a classical Big Bounce. 
\end{abstract}

\section*{Introduction}

Vector fields play a prominent role in modern particle physics. We detect vector fields in laboratory and we expect them to participate actively in the physics of the early Universe. As shown in literature, vector fields can contribute to generation of the observed large scale structure \cite{Dimopoulos:2006ms,Dimopoulos:2008yv,Watanabe:2010fh,Gumrukcuoglu:2010yc,Dulaney:2010sq}, background anisotropies \cite{Koivisto:2008xf}, inflation \cite{Golovnev:2008cf} and gravitational waves \cite{Golovnev:2008hv}, so that, in principle,  one could reconstruct standard FRW cosmology without scalar fields. In order to obtain  slow-roll evolution and flat power spectra of initial inhomogeneities, vector fields need to have a time dependent mass term. It may originate from the Higgs mechanism, non-canonical kinetic terms or from a non-minimal coupling to gravity.

On the other hand, a study of the influence of Loop Quantum Cosmology correction on evolution of FRW universes \cite{Ashtekar:2006wn,Bojowald:2006da,Szulc:2006ep,Ashtekar:2006es,Artymowski:2008sc} and  Bianchi I models \cite{Szulc:2008ar,Chiou:2007dn,Chiou:2006qq,Ashtekar:2009vc,MartinBenito:2009qu} has been performed. In both cases the existence of the quantum of length changes significantly the evolution of scale factors near the Planck scale, resolving the problem of the initial singularity. 

The Big Bounce model usually assumes, that the Universe is filled with a homogeneous scalar field, so there is no physical argument for breaking the $O(3)$ symmetry of the metric tensor.  In this paper we shall analyse the LQC in the Universe filled with a vector field. We start with a simple example, where the Universe is dominated by a massless or a massive vector field. Next, to study the influence of the vector field in more realistic set-ups we consider a case, where  a vector and a scalar fields are present. They can be uncoupled, or strongly coupled. The last case corresponds in this paper to the time dependent mass term for the vector field.  An interesting  idea is to introduce vectors with a non-minimal coupling to gravity, however it is not known how to take into account the LQC corrections in this case. 

In all of those scenarios the vector field produces anisotropic pressure, which forms a physical source for the anisotropic expansion in the Bianchi I model. The main purpose of this paper is to find out, whether vectors could change significantly the behaviour of scale factors close to the moment of the Big Bounce. 

In this paper we investigate two generalisations of the isotropic LQC model. Two schemes of constructing the $\bar{\mu}_i$ functions are presented and compared. It is worth to note, that both of them give similar results. 
\\*

This paper is organized as follows: In section \ref{sec:rozdzial1} we introduce the Hamiltonian formalism for the metric tensor with two scale factors and we calculate the Einstein's equation together with equations of motion for matter fields. In section \ref{sec:rozdzial2} we analyse the influence of the loop correction on the classical evolution of scale factors and matter fields for the case of the vector field domination. In section \ref{sec:rozdzial3} we study the Universe dominated by both,  vector and scalar fields in the context of LQC. Conclusions are presented in section \ref{sec:concl}

\section{The classical theory}\label{sec:rozdzial1}
\subsection{Ashtekar formalism and equations of motion}
In this section we introduce Ashtekar variables for the anisotropic Universe. The full discussion of this issue can be found in \cite{Szulc:2008ar,Manojlovic:1991dk}. Let us consider a homogeneous, massive vector field \cite{Dimopoulos:2006ms} $A_\mu=A(t)\delta^3_{\ \mu}$, which contributes to the Universe's energy density. Since the vector field breaks the $O(3)$ symmetry of the Universe, one needs to introduce two scale factors - $b(t)$ for the direction pointed by the vector and $a(t)$ for transverse directions. This means, that the model is a simplified version of diagonal Bianchi I. Let us set the lapse function to be $N(t)=1$, which means, that the metric tensor is given by
\begin{equation}
ds^2=dt^2-a^2(t)(dx^2+dy^2)-b^2(t)dz^2\ \Rightarrow g_{\mu\nu}=Diag(1,-a^2,-a^2,-b^2)\ . \label{eq:metryka} 
\end{equation}
Using the eq. (\ref{eq:metryka}) one can introduce the canonically conjugated Ashtekar variables \cite{Szulc:2008ar}
\begin{equation}
p_1=L_1L_2ab\ ,\qquad p_2=L_1^2a^2\ ,\qquad c_1=L_1\gamma\dot{a}\ , \qquad c_2=L_2\gamma\dot{b}\ , \label{eq:zmienneAshtekara}
\end{equation}
where $\gamma$ is the Barbero-Immirzi parameter and $\{c_1,p_1 \}= \frac{8 \pi G \gamma} {2}$, $\{c_2,p_2 \}= 8 \pi G \gamma $. $L_1,L_2$ are comoving lengths introduced to make Ashtekar variables independent of the rescaling of scale factors. One constructs the fiducial cell $V_o=L_1^2L_2$, which is used to obtain finite scalar constraint. Ashtekar variables $p_i$ have the following interpretation: the area perpendicular to the $z$ axis is proportional to $a^2$ and areas perpendicular to $x$ an $y$ axes are proportional to $ab$. On the other hand variables $c_1$ and $c_2$ have the interpretation of external curvature along $x,y$ and $z$ axes respectively. Together with Ashtekar variables we shall also define the total physical volume of the fiducial cell $V=(p_1^2p_2)^{1/2}$. For the metric tensor described by (\ref{eq:metryka}) one obtains $V\propto a^2b$, so $V$ is proportional to the volume of the Universe. Let us assume, that the Universe is filled with a vector field and a scalar field\footnote{In stead of a scalar field one can consider a perfect fluid.}.  With the convention $8\pi G=1$ the scalar constraint for the minimally coupled general relativity looks as follows
\begin{equation}
\mathcal{C}=-\frac{1}{\gamma^2p_1\sqrt{p_2}}(c_1^2p_1^2+2c_1c_2p_1p_2)+\frac{p_1}{2p_2^{3/2}}\pi_A^2+\frac{p_2^{3/2}}{2p_1}m^2A^2+\frac{1}{2p_1\sqrt{p_2}}\pi^2_\phi+p_1\sqrt{p_2}V(\phi)=0\ ,\label{eq:Hamiltonian}
\end{equation}
where $\pi_A=p_2^{3/2}\dot{A}/p_1=a^2\dot{A}/b$ and $\pi_\phi=p_1\sqrt{p_2}$ are canonical momenta of vector and scalar fields. The scalar constraint decomposes into $\mathcal{C}=\mathcal{C}_{grav}+\mathcal{C}_{mat}$, where $\mathcal{C}_{mat}=p_1\sqrt{p_2}\rho$. Equations of motion for the Ashtekar variables are given by Hamilton equations \footnote{Coefficients $\gamma$ and $\gamma/2$ come from the symmetry reduction of diagonal Bianchi I model. They appear because $\{c_1,p_1\}=\gamma/2$ and $\{c_2,p_2\}=\gamma$.}
\begin{eqnarray}
\dot{p}_1=\{p_1,\mathcal{C}\}=-\frac{\gamma}{2}\frac{\partial \mathcal{C}}{\partial c_1}\ ,\qquad\dot{p}_2=
\{p_2,\mathcal{C}\}=-\gamma\frac{\partial \mathcal{C}}{\partial c_2}\ ,\label{eq:ruchupi} \\
\dot{c}_1=\{c_1,\mathcal{C}\}=\frac{\gamma}{2}\frac{\partial \mathcal{C}}{\partial p_1}\ ,\qquad\dot{c}_2=\{c_2,\mathcal{C}\}=\gamma\frac{\partial \mathcal{C}}{\partial p_2}\ , \label{eq:ruchuci}
\end{eqnarray}
which together with the eq. (\ref{eq:Hamiltonian})  give Einstein's equations and the equations of motion for the matter fields\footnote{We refer to scalar and vector fields collectively as matter fields.}
\begin{eqnarray}
H(H+2\mathcal{H})=\rho=\rho_\phi+\rho_A\ , \label{eq:Ein00}\\
\frac{\ddot{a}}{a}+\frac{\ddot{b}}{b}+H\mathcal{H}=-p_\perp=-p_\phi-p_A\ , \label{eq:Ein11}\\
2\frac{\ddot{a}}{a}+H^2=-p_\parallel=-p_\phi+p_A\ , \label{eq:Ein33}\\
\dot{\rho}_\phi+(2H+\mathcal{H})(\rho_\phi+p_\phi)=0\ ,\qquad \dot{\rho}_A+2H(\rho_A+p_A)+\mathcal{H}(\rho_A-p_A)=0\ , \label{eq:ciaglosci}
\end{eqnarray}
where $H=\frac{\dot{a}}{a}\ ,\mathcal{H}=\frac{\dot{b}}{b}$. These equations, together with equations of motion for matter fields, give the complete evolution of the classical space-time. In realistic cosmological models one expects the vector field to produce small background anisotropies of the Universe, so one could expect $\lambda=H-\mathcal{H}\ll H+\mathcal{H}$. Nonetheless, as we shall see later the vector field domination leads to observable anisotropies.

Let us focus for a while on the background evolution of the vector field. One can show, that
\begin{equation}
\rho_A=\frac{1}{2b^2}(\dot{A}^2+m^2A^2),\qquad p_A=\frac{1}{2b^2}(\dot{A}^2-m^2A^2).\label{eq:rhoApA}
\end{equation}
The pressure and the energy density are physical quantities, so they are invariant under the re-scaling of any scale factor. This means, that $A$ transforms like $A\rightarrow l_zA$ under the rescaling $b\rightarrow l_zb$. Then $A(t)$ shall be replaced by the physical field $U(t)=A/b$, which gives $\frac{1}{b^2}A^2=U^2$ and $\frac{1}{b^2}\dot{A}^2=\dot{U}^2+2\mathcal{H}\dot{U}U+\mathcal{H}^2U^2$. The eq. (\ref{eq:ciaglosci}) looks now as follows
\begin{equation}
\ddot{U}+(2H+\mathcal{H})\dot{U}+(2\mathcal{H}H+\dot{\mathcal{H}}+m^2)U=0.\label{eq:ruchuU}
\end{equation}
One can see, that $U$ has time-dependent effective mass term $\tilde{m}^2=2\mathcal{H}H+\dot{\mathcal{H}}+m^2$. For $a=b$ one obtains $\tilde{m}^2=2H^2+\dot{H}+m^2=R/6+m^2$, so $U$ shall be very heavy in the high energy regime, in which $H^2\gg m^2$. During inflation the subdominant $U$ is much heavier than the inflaton, so after a few e-folds it reaches its minimum and starts oscillating. The energy density of a vector field behaves matter-like, $\rho_A\propto a^{-3}\propto e^{-3Ht}$, and vectors are expected to be washed out during inflation. On the other hand the power spectrum of the vector field perturbation would not be flat enough to fit the CMB data. For the time of the cosmic inflation we need thus an extra tachyonic mass term $m^2\simeq -2H^2$ which would allow the slow roll evolution. After the reheating the vector field  could play a role of the curvaton and dominate the Universe via its oscillations. One can learn more about the issue of the vector field effective mass from e.g.  \cite{Dimopoulos:2006ms,Dimopoulos:2008yv,Golovnev:2008cf}. 

One shall remember, that the Ashtekar formalism gives us the same results as the Lagrange formalism with the action
\begin{equation}
S=\int d^4x\sqrt{-g}\left(\frac{1}{2}R-\frac{1}{4}F_{\mu\nu}F^{\mu\nu}+\frac{1}{2}m^2A^\mu A_\mu\right) \ , \label{eq:dzialanieA}
\end{equation}
where $R$ is the Ricci scalar of the anisotropic Universe
\begin{equation}
R=2(H^2+2H\mathcal{H}+2\frac{\ddot{a}}{a}+\frac{\ddot{b}}{b})\ . \label{eq:krzywizna}
\end{equation}

\subsection{Massless vector field domination} \label{sec:bezmasowywektorKLAS}
The most popular case in the LQC is a scalar field without any potential term. This produces simple solution and allows a matter field to play a role of a time variable. Since it is the most fundamental calculation in LQC we will first of all consider this case for the vector field in the GR frame. As we will see later on this model will bring us rather non physical solutions of Einstein equations. On the other hand those solutions shall describe the low energy limit for the massless vector field in the LQC frame. When the energy-stress tensor consist only of the massless vector field, then $\rho_A=\rho=p$. From eq. (\ref{eq:Ein00},\ref{eq:Ein33}) one obtains
\begin{equation}
H^2+2H\mathcal{H}=2\frac{\ddot{a}}{a}+H^2\Rightarrow \frac{\ddot{a}}{\dot{a}}=\frac{\dot{b}}{b}\Rightarrow b=E\dot{a}\ , \label{eq:b=aprim}
\end{equation}
where $E=const>0$ has a dimension of time. This shows, that the Universe shrinks along the $z$ axe if only $\ddot{a}<0$. On the other hand, the eq. (\ref{eq:ciaglosci}) gives
\begin{equation}
\dot{\rho}+4H\rho=0\Rightarrow \rho=\rho_o\left(\frac{a_o}{a}\right)^4\ , \label{eq:rhooda}
\end{equation}
where $\rho_o=\rho(t_o)$ and $t_o$ is any fixed moment in the Universes history. This equation can be also obtained from the equation of motion of the vector field. For the massless vector field one gets
\begin{equation}
\ddot{A}+(2H-\mathcal{H})\dot{A}=0\Rightarrow \dot{A}=\exp\left[\int(\mathcal{H}-2H)dt\right] \ . \label{eq:AdotodH}
\end{equation} 
Then, from the eq. (\ref{eq:b=aprim}) one obtains
\begin{equation}
\dot{A}=D\frac{a_o}{a}H\ ,\Rightarrow \rho=\frac{D^2a_o^2}{2a^2b^2}H^2=\frac{D^2a_o^2}{2E^2a^4}\ , \label{eq:Adotoda}
\end{equation}
where $D=const>0$. From the eq. (\ref{eq:rhooda}) one obtains $D=\sqrt{2\rho_o}a_oE$. To calculate $a(t)$ let us combine eq. (\ref{eq:Ein33},\ref{eq:rhooda}) which together give
\begin{equation}
2\ddot{a}a+\dot{a}^2=\rho_oa_o^4a^{-2}\ . \label{eq:ruchua}
\end{equation}
This equation can be simplified to
\begin{equation}
\dot{a}=a_o\sqrt{C_1\frac{a_o}{a}-\rho_o\frac{a_o^2}{a^2}} \ ,\label{adotoda}
\end{equation}
where $C_1=const$ has a dimension of energy. One can see, that this solution faces a constraint. Let us define the initial value of the scale factor $a$ by $a_I=a(t_I)=a_oC_1/\rho_o$. Then $\dot{a}(a_I)=b(a_I)=0$, which means, that to obtain real values of $\dot{a},b$ one needs $a\geq a_I$. Thus the energy density cannot be greater than $\rho_I=\rho_o(a_o/a_I)^4$. It is important to note, that the initial (maximal) energy density is simply a constant of integration. So far let us make no assumptions about $\rho_I$, but later on we shall investigate a massless vector in the LQC frame, for which the natural scale of the maximal energy density is the Planck scale. Thus, $\rho_I$ will set by hand to that scale. Formulae presented so far take the simplest form for $t_o=t_I$, which is the choice we shall adopt.

\begin{figure}[h]
\includegraphics*[height=10.5cm]{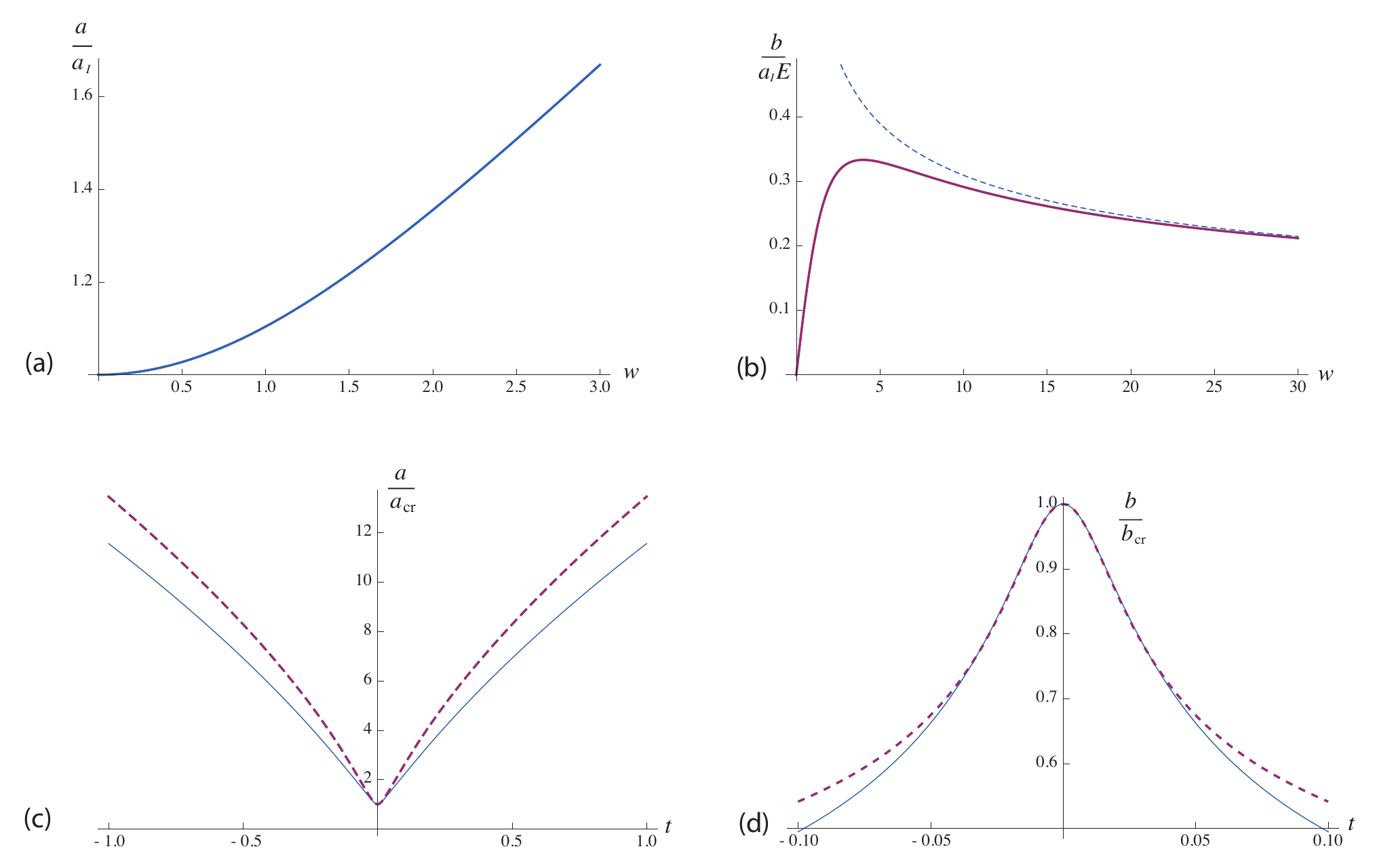}
\caption{In the panel $(a)$ we present the evolution of $a(w)/a_I$ for massless vector field domination in GR. Panel $(b)$ shows the evolution of $b(w)/(a_I E)$. The red line corresponds to $b(w)$, and the blue dashed line corresponds to $b(w)\propto w^{-1/3}$, which is the low energy limit for $b(w)$ evolution. Panels $(c)$ and $(d)$ show $a(t)/a_{cr}$ and $b(t)/b_{cr}$ as a function of time in Planck units for the LQC with the massless vector field domination. Blue and red dashed lines corresponds to the old and the new $\bar{\mu}_i$ scheme respectively. One can see, that $b_{cr}$ is the global maximum of $b(t)$.}
\label{fig:abklas-kwant}
\end{figure}

So far we have calculated $H,\mathcal{H},b,\rho,\dot{A}$ as functions of the scale factor $a$. Thus, to find time dependence of those quantities one needs to find a solution of the eq. (\ref{eq:ruchua}), which is 
\begin{equation}
\left(\frac{a}{a_I}\right)^3+3\left(\frac{a}{a_I}\right)^2-4=w^2=\frac{9}{4}\rho_I(t-t_I)^2\ . \label{eq:aodt}
\end{equation}
By shifting the time coordinate $t\rightarrow t+t_I$ one obtains $w=\frac{3}{2}\sqrt{\rho_I}t$. The exact solution of (\ref{eq:aodt}) is
\begin{equation}
a(t)=a_I\left[\frac{2^{1/3}}{(2+w^2+\sqrt{4w^2+w^4})^{1/3}}+\frac{(2+w^2+\sqrt{4w^2+w^4})^{1/3}}{2^{1/3}}-1\right]\ . \label{eq:aodtdokladnie}
\end{equation}
From eq. (\ref{eq:krzywizna},\ref{eq:b=aprim},\ref{eq:aodtdokladnie}) one obtains $R=0$ for any $t$. For $t\to 0$ one also obtains $R^{\mu\nu}R_{\mu\nu}\to 4\rho_I^2$ and $R^{\mu\nu\alpha\beta}R_{\mu\nu\alpha\beta}\to 20\rho_I^2$. Thus the massless vector field domination gives no initial curvature singularity of the Universe \footnote{The energy conditions for the vector field domination require $\rho>0$, $p>0$ and $\rho\geq|p|$. The massless vector field satisfies them, since $p=\rho=\dot{A}^2/2b^2>0$.}, which together with $a_I\neq 0, H(t_I)=0$ are features of the Big-Bounce. On the other hand from (\ref{eq:b=aprim},\ref{eq:aodt}) one obtains $\mathcal{H}\rightarrow\infty, b\rightarrow 0$ for $t\rightarrow 0$, which looks alike to the Big Bang scenario. When we are very close to the initial value of the energy density one obtains $w\ll 1$, which gives
\begin{equation}
a\simeq a_I\left(1+\frac{1}{9}w^2\right)=a_I\left(1+\frac{1}{4}\rho_It^2\right)\ , \qquad b\propto t \ ,\qquad V\propto t\left(1+\frac{1}{4}\rho_It^2\right)^2\ . \label{eq:a,bodtdlaw<<1}
\end{equation}
The Universe expands along the $z$ direction for $w\in (0,4)$. Let us note, that $a(t)$ for $w\ll1$ evolves like $a(t)$ around the Big Bounce in the dust dominated isotropic Universe in LQC frame \footnote{Let us note, that the Big Bounce in the GR frame may be also obtained in i.e. cycling Universe \cite{Steinhardt:2001st} and $f(R)$ gravity \cite{Gurovich}.}. For $\rho\ll\rho_I$ one obtains $w\gg 1$. Then
\begin{equation}
a\simeq a_I w^{2/3}=a_I\left(\frac{9}{4}\rho_I\right)^{1/3}t^{2/3}\ ,\qquad b\propto t^{-1/3}\propto a^{-1/2} \ ,\qquad V\propto t\ . \label{eq:a,bodtdlaw>>1}
\end{equation}
The low energy limit bring us to the Kasner-like solution of Bianchi I model, with one direction shrinking and two expanding. The model with such predictions is obviously excluded by the astronomical observations, but as we have mentioned it is just a GR limit of the Universe with the LQC correction, which shall be discussed later. The evolution of $a(w)$ and $b(w)$ is shown in the fig. \ref{fig:abklas-kwant}
\\*

\subsection{Massive vector field}\label{sec:masywnywektorKLAS}
\begin{figure}[h]
\centering
\includegraphics*[height=5cm]{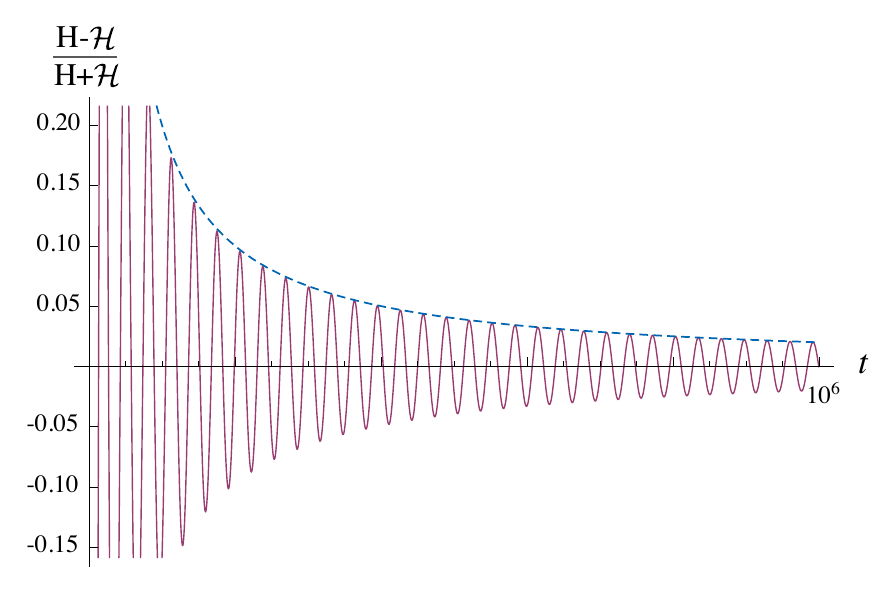}
\hspace{0.5cm}
\includegraphics*[height=5cm]{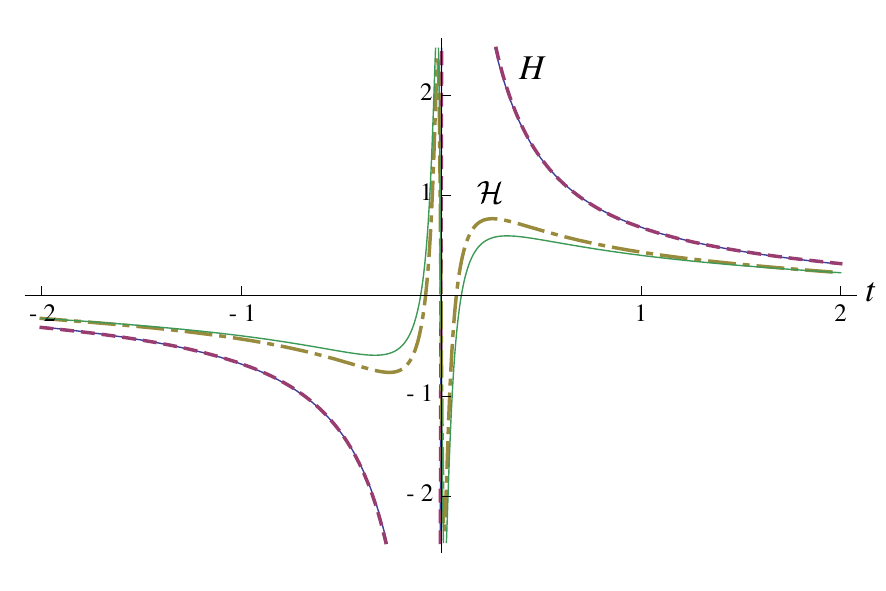}
\caption{In the right panel we present the evolution of $H$ (blue and red lines) and $\mathcal{H}$ (yellow and green lines) for domination of a scalar and a vector field with a strong coupling in the LQC frame. The old and the new $\bar{\mu}_i$ scheme are denoted as dashed and continuous lines respectively.  For the numerical calculations we have assumed purely kinetic initial conditions. In the left panel we present the evolution of the Universe anisotropy during the period of the domination of the oscillating vector field. Dashed line denotes the analytical solution for $(H-\mathcal{H})/(H+\mathcal{H})$ for the pressureless anisotropic Universe. This solution corresponds to $\rho\ll\rho_{cr}$, where both schemes of $\bar{\mu}_i$ gives the same results}
\label{fig:HiggsHaHb}
\end{figure}

In this subsection we shall investigate the case of a massive vector field domination with $m=const$. In our calculations in GR frame we wanted to construct the low-energy limit for the LQC solutions. Since then we are interested in the massive vector field with purely kinetic initial conditions similar to the massless field scenario \ref{sec:bezmasowywektorKLAS}. If initially $A_I=0$ then we can describe the beginning of a Universe by the massless vector field domination. Thus, from the eq. (\ref{eq:Adotoda}) one obtains
\begin{equation}
A=\sqrt{2\rho_I}Ea_I\left(1-\frac{a_I}{a}\right)\simeq\frac{1}{4}\sqrt{2\rho_I}Ea_I\rho_It^2,\quad U=\sqrt{2\rho_I}\frac{a_I}{\dot{a}}\left(1-\frac{a_I}{a}\right)\label{eq:Aodtmass}
\end{equation}
After a short period of this analytically described evolution we need to take into account corrections form the mass term. Then, for the numerical simulation we shall assume a finite $\mathcal{H}(t_\star)$and a non-zero $H(t_\star)$, where $\star$ denotes the boundary condition for analytical and initial condition for numerical calculation. Regardless of any initial conditions one obtains $\mathcal{H}>0$ for e.g. $t>O(10^3t_{pl})$ and $m=10^{-4}$. When the energy density density decreases, and $m>\max\{H,\mathcal{H}\}$, the vector field starts to oscillate and one obtains the matter-like dominated Universe. The $A_{\mu}$ becomes pressureless, so there are no anisotropies in the energy-stress tensor. Then, as it is shown in the fig. \ref{fig:HiggsHaHb}, the anisotropy of the Universe defined by $(H-\mathcal{H})/(H+\mathcal{H})$ decreases like $t^{-1}$ .
\\*

\section{Semiclassical evolution in LQC}\label{sec:rozdzial2}
\subsection{Holonomy correction to Ashtekar variables}\label{sec:ruchuLQC}

As shown in the previous section, the massless vector field domination gives some predictions similar to the LQC. Nevertheless, it is worth to investigate the impact of vector fields on the LQC diagonal Bianchi I model - we shall compare it with results from GR and investigate more realistic models inspired by cosmic inflation. In this section we shall analyse the influence of the LQC holonomy correction on the classical evolution of scale factors, scalar field and a vector field. We shall focus on the effective, semi-classical equations of motion. Thus, we treat LQC as a modified, but still classical theory of gravity, alike $f(R)$ theory. This means, that some effects, which come from a quantum nature of LQC, may be missing. We disregard the inverse triad correction, so only $c_1$ and $c_2$ are modified 
\begin{equation}
c_i\rightarrow \frac{1}{\bar{\mu}_i}\sin\left(\bar{\mu}_i c_i\right)\ ,
\end{equation}
where $\bar{\mu}_i=\bar{\mu}_i(p_1,p_2)$ will be specified later on. The effective gravitational Hamiltonian is now given by
\begin{equation}
\mathcal{C}_{eff}=-\frac{1}{\gamma^2p_1\sqrt{p_2}}\left[\sin^2\left(\bar{\mu}_1 c_1\right)\frac{p_1^2}{\bar{\mu}_1^2}+2\sin\left(\bar{\mu}_1 c_1\right)\sin\left(\bar{\mu}_2 c_2\right)\frac{p_1p_2}{\bar{\mu}_1\bar{\mu}_2}\right]\ . \label{eq:Ceff}
\end{equation} 
One shall remember, that the loop correction changes the physical interpretation of $c_i$. From now on $c_1\neq \gamma \dot{a}$ and $c_2\neq\gamma\dot{b}$. The low energy limit of the quantum theory shall give the General Relativity. Indeed, for the $\bar{\mu}_i c_i\ll 1$ one obtains
\begin{equation}
\frac{1}{\bar{\mu}_i}\sin\left(\bar{\mu}_i c_i\right)\rightarrow c_i\ , \qquad \mathcal{C}_{eff}\rightarrow \mathcal{C}_{grav}. \label{eq:LQCniskoenergetyczna}
\end{equation}
The $\mathcal{C}_{mat}$ remains unchanged under the loop correction, since it is independent of $c_i$. The equation $\mathcal{C}=\mathcal{C}_{grav}+\mathcal{C}_{mat}=0$ now takes the form $\mathcal{C}=\mathcal{C}_{eff}+\mathcal{C}_{mat}=0$.

\subsection{The old $\bar{\mu}$ scheme}\label{sec:stary}

In the isotropic LQC model one defines $\bar{\mu}$ as $\sqrt{\Delta/p}$. To analyse the Bianchi I model one needs to introduce a generalisation of $\bar{\mu}$. There are two models of LQC Bianchi I discussed in literature. The first one, described by e.g. \cite{Szulc:2008ar} defines $\bar{\mu}_i$ as $\bar{\mu}_i=\sqrt{\Delta/p_i}$, where $\Delta$ is the eigenvalue of the area operator \cite{Ashtekar:1996eg}. From the scalar constraint one obtains 
\begin{equation}
\rho=\frac{1}{\Delta\gamma^2}\left[\sin^2\left(\frac{\sqrt{\Delta}c_1}{\sqrt{p_1}}\right)\frac{p_1}{p_2}+2\sin\left(\frac{\sqrt{\Delta}c_1}{\sqrt{p_1}}\right)\sin\left(\frac{\sqrt{\Delta}c_2}{\sqrt{p_2}}\right)\sqrt{\frac{p_2}{p_1}}\right]\ . \label{eq:rhopetla}
\end{equation}
Equations (\ref{eq:ruchupi},\ref{eq:ruchuci}) are unchanged under the loop correction and since $\dot{p}_1=ab(H+\mathcal{H})$ and $\dot{p}_2=2a^2H$ they give the Hubble parameters
\begin{eqnarray}
H=\frac{1}{\sqrt{\Delta}\gamma}\sin\left(\frac{\sqrt{\Delta}c_1}{\sqrt{p_1}}\right)\cos\left(\frac{\sqrt{\Delta}c_2}{\sqrt{p_2}}\right)\sqrt{\frac{p_1}{p_2}} \ ,\qquad\qquad\qquad\qquad\qquad\label{eq:Ha}\\
\mathcal{H}=\frac{1}{\sqrt{\Delta}\gamma}\left[\sin\left(\frac{\sqrt{\Delta}c_1}{\sqrt{p_1}}\right)\left(\cos\left(\frac{\sqrt{\Delta}c_1}{\sqrt{p_1}}\right)-\cos\left(\frac{\sqrt{\Delta}c_2}{\sqrt{p_2}}\right)\right)\sqrt{\frac{p_1}{p_2}}+\sin\left(\frac{\sqrt{\Delta}c_2}{\sqrt{p_2}}\right)\cos\left(\frac{\sqrt{\Delta}c_1}{\sqrt{p_1}}\right)\frac{p_1}{p_2}\right]. \label{eq:Hb}
\end{eqnarray}
To obtain the Big Bounce we need $H=\mathcal{H}=0$ at the Planck scale. There are two limiting cases, which are of interest: the low energy limit which we have already discussed, and the Planck energy limit with significant corrections from LQC. In the latter case it is easy to check, that
\begin{equation}
\sin\left(\frac{\sqrt{\Delta}c_1}{\sqrt{p_1}}\right)=\sin\left(\frac{\sqrt{\Delta}c_2}{\sqrt{p_2}}\right)=1\Rightarrow H=\mathcal{H}=0\ . \label{eq:warunkiBB}
\end{equation}
The critical energy density which corresponds to the Big Bounce is
\begin{equation}
\rho_{cr}=\frac{1}{\Delta\gamma^2}\left(\frac{L_2b_{cr}}{L_1a_{cr}}+2\sqrt{\frac{L_1a_{cr}}{L_2b_{cr}}}\right)\ , \label{eq:rhocrodab}
\end{equation}
where $a_{cr}$ and $b_{cr}$ are the values of scale factors during the Big Bounce. One obtains the minimal value of the critical energy density for $L_1a_{cr}=L_2b_{cr}$, which is $\rho_{cr}=\frac{3}{\Delta\gamma^2}$\footnote{There are two values of $\rho_{cr}$, which are often considered in literature: $\rho_{cr}\simeq 0.82G^2$ and $\rho_{cr}\simeq 0.41 G^2$. To find more about this issue see \cite{Ashtekar:2009vc}. For all numerical calculations we have chosen the $\rho_{cr}\simeq 0.82G^2$, which means, that $\Delta\simeq3/(\gamma^2 0.82G^2)$}. The choice of the physical length scales related to values of $p_1(t_{cr})$ and $p_2(t_{cr})$ is arbitrary, for simplicity we set them to be equal. Unfortunately other choices give different evolution of  $p_(t)$ and $p_2(t)$, which is a problem of anisotropic models of LQC in the old $\bar{\mu}_i$ scheme. The dependence of Bianchi I evolution on the choice of the fiducial cell and its rescaling is discussed in e.g. \cite{Szulc:2006ep,Ashtekar:2009vc}. 

Let us study effective equations of motion for $c_1$ and $c_2$. From eq. (\ref{eq:ruchuci}) one obtains
\begin{eqnarray}
\dot{c}_1=\frac{1}{2\Delta\gamma}\Bigg[\sin\left(\frac{\sqrt{\Delta}c_1}{\sqrt{p_1}}\right)\cos\left(\frac{\sqrt{\Delta}c_1}{\sqrt{p_1}}\right)\sqrt{\frac{p_1}{p_2}}\sqrt{\Delta}c_1-2\sin^2\left(\frac{\sqrt{\Delta}c_1}{\sqrt{p_1}}\right)\frac{p_1}{\sqrt{p_2}}+\nonumber\\ \cos\left(\frac{\sqrt{\Delta}c_1}{\sqrt{p_1}}\right)\sin\left(\frac{\sqrt{\Delta}c_2}{\sqrt{p_2}}\right)\frac{p_2}{p_1}\sqrt{\Delta}c_1-\sin\left(\frac{\sqrt{\Delta}c_1}{\sqrt{p_1}}\right)\sin\left(\frac{\sqrt{\Delta}c_2}{\sqrt{p_2}}\right)\frac{p_2}{\sqrt{p_1}}\Bigg]+\frac{\gamma}{2}(p_A-p_\phi)\sqrt{p_2}\ ,\label{eq:c1pot}\\
\dot{c}_2=\frac{1}{\Delta\gamma}\Bigg[\frac{1}{2}\sin^2\left(\frac{\sqrt{\Delta}c_1}{\sqrt{p_1}}\right)\frac{p_1^2}{p_2^{3/2}}+\sin\left(\frac{\sqrt{\Delta}c_1}{\sqrt{p_1}}\right)\cos\left(\frac{\sqrt{\Delta}c_2}{\sqrt{p_2}}\right)\sqrt{\frac{p_1}{p_2}}\sqrt{\Delta}c_2-\nonumber\\
2\sin\left(\frac{\sqrt{\Delta}c_1}{\sqrt{p_1}}\right)\sin\left(\frac{\sqrt{\Delta}c_2}{\sqrt{p_2}}\right)\sqrt{p_1}\Bigg]-\frac{p_1}{2\sqrt{p_2}}\gamma (3p_A+p_\phi)\ . \label{eq:c2pot}
\end{eqnarray}
Those equations together with eq. (\ref{eq:Ha},\ref{eq:Hb}) and equations of motion for matter fields give us the complete dynamics of the Universe.

\subsection{The new $\bar{\mu}$ scheme}\label{sec:nowy}

The problem of arbitrariness of the scaling in LQC of Bianchi I model may be solved by the redefinition of the $\bar{\mu}_i$. Let us consider $\bar{\mu}_1=\sqrt{\Delta/p_2}$, $\bar{\mu}_2=\sqrt{\Delta p_2}/p_1$ \cite{Chiou:2007sp,Ashtekar:2009vc}. Then one obtains $\rho_{cr}=\frac{3}{\Delta\gamma^2}$, so the critical energy density does not depend on the size of the fiducial cell. Form the Hamilton equations one finds
\begin{eqnarray}
H=\frac{1}{\gamma\sqrt{\Delta}}\sin\left(\frac{\sqrt{\Delta}c_1}{\sqrt{p_2}}\right)\cos\left(\frac{\sqrt{\Delta}c_2\sqrt{p_2}}{p_1}\right), \qquad\qquad\qquad\label{eq:HLQCnowy}\\
\mathcal{H}=\frac{1}{\gamma\sqrt{\Delta}}\cos\left(\frac{\sqrt{\Delta}c_1}{\sqrt{p_2}}\right)\left[\sin\left(\frac{\sqrt{\Delta}c_1}{\sqrt{p_2}}\right)+\sin\left(\frac{\sqrt{\Delta}c_2\sqrt{p_2}}{p_1}\right)\right]-H, \label{eq:hLQCnowy}\qquad\\
\dot{c}_1=\frac{1}{\gamma\sqrt{\Delta}}\sqrt{p_2}\sin\left(\frac{\sqrt{\Delta}c_1}{\sqrt{p_2}}\right)\Bigg[\sqrt{\Delta}c_2\frac{\sqrt{p_2}}{p_1}\cos\left(\frac{\sqrt{\Delta}c_2\sqrt{p_2}}{p_1}\right)-\sin\left(\frac{\sqrt{\Delta}c_2\sqrt{p_2}}{p_1}\right)-\nonumber\\
\frac{1}{2}\sin\left(\frac{\sqrt{\Delta}c_1}{\sqrt{p_2}}\right)\Bigg]+\frac{\gamma}{2}(p_A-p_\phi)\sqrt{p_2}, \qquad\qquad\qquad\quad\label{eq:c1potnowy}\\
\dot{c}_2=-\frac{p_1}{2\sqrt{p_2}}\gamma (3p_A+p_\phi)-\frac{1}{\gamma\sqrt{\Delta}}\frac{p_1}{\sqrt{p_2}}\Bigg[\frac{1}{2}\sin^2\left(\frac{\sqrt{\Delta}c_1}{\sqrt{p_2}}\right)+\sin\left(\frac{\sqrt{\Delta}c_1}{\sqrt{p_2}}\right)\cos\left(\frac{\sqrt{\Delta}c_1}{\sqrt{p_2}}\right)+\nonumber\\
\sqrt{\Delta}c_2\frac{\sqrt{p_2}}{p_1}\sin\left(\frac{\sqrt{\Delta}c_1}{\sqrt{p_2}}\right)\cos\left(\frac{\sqrt{\Delta}c_2\sqrt{p_2}}{p_1}\right)-\frac{\sqrt{\Delta}c_1}{\sqrt{p_2}}\cos\left(\frac{\sqrt{\Delta}c_1}{\sqrt{p_2}}\right)\sin\left(\frac{\sqrt{\Delta}c_2\sqrt{p_2}}{p_1}\right)\Bigg].\label{eq:c2potnowy}
\end{eqnarray}
One can see, that this choice of $\bar{\mu}_i$ functions changes not only the dependence on the size of the fiducial cell, but also dynamics of Ashtekar variables. Nevertheless, as we will show, evolution of matter field and scale factors are similar in both schemes. From the scalar constraint and eq. (\ref{eq:HLQCnowy},\ref{eq:hLQCnowy}) one obtains
\begin{eqnarray}
\mathcal{H}=-H+\frac{1}{3 \rho ^2+4 H^2 \rho _{cr}}\left(3 \rho ^2+\left(2 H^2+\rho +\rho  \sqrt{1-\frac{3 \left(H^2-\rho \right)}{\rho _{cr}}}\right) \rho _{cr}\right)\times\nonumber\\
 \sqrt{\frac{-3 \rho ^2}{\rho _{cr}}+\left(4 H^2+\rho  \left(4 \sqrt{1-\frac{3 \left(H^2-\rho \right)}{\rho _{cr}}}-7\right)+2 \left(\sqrt{1-\frac{3 \left(H^2-\rho\right)}{\rho _{cr}}}-1\right) \rho _{cr}\right)},
\end{eqnarray}
which is the analytical relation between all measurable quantities of the model. It is worth to note, that for $\rho\simeq\rho_{cr}$ one obtains $\mathcal{H}\simeq -H$, which is confirmed by numerical results. 

Let us discuss several scenarios for the vector field contribution to the evolution of the Universe. Usually one considers the isotropic Universe with subdominant vector field during inflation or radiation domination, or with the vector field domination (or its significant contribution) during the vector field oscillation phase. In following sections we shall analyse extensions of those model to the anisotropic Universe with the vector field domination and LQC corrections for both $\bar{\mu}_i$ schemes.
\\*

\subsection{Massless vector field domination}\label{sec:AbezmasLQC}

\begin{figure}[h]
\centering
\includegraphics*[height=3.6cm]{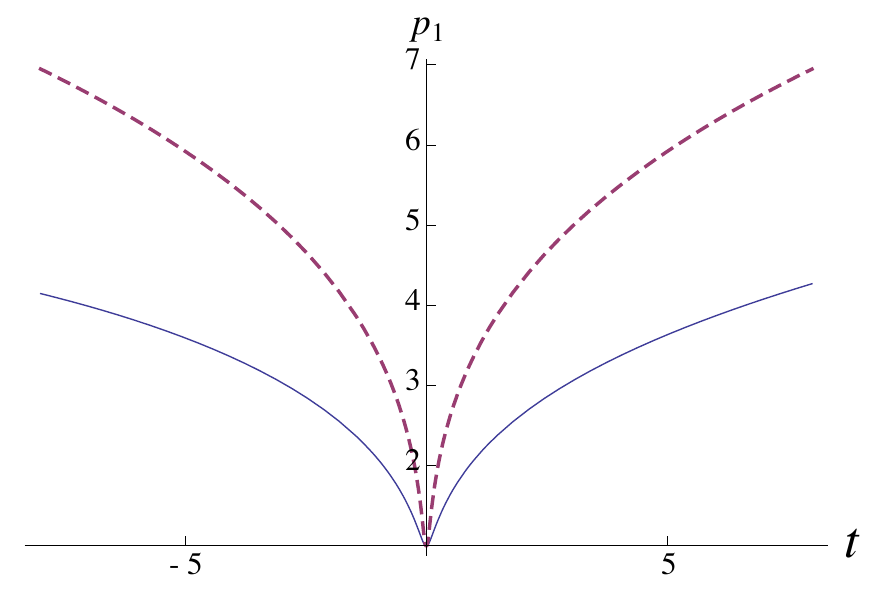}
\includegraphics*[height=3.6cm]{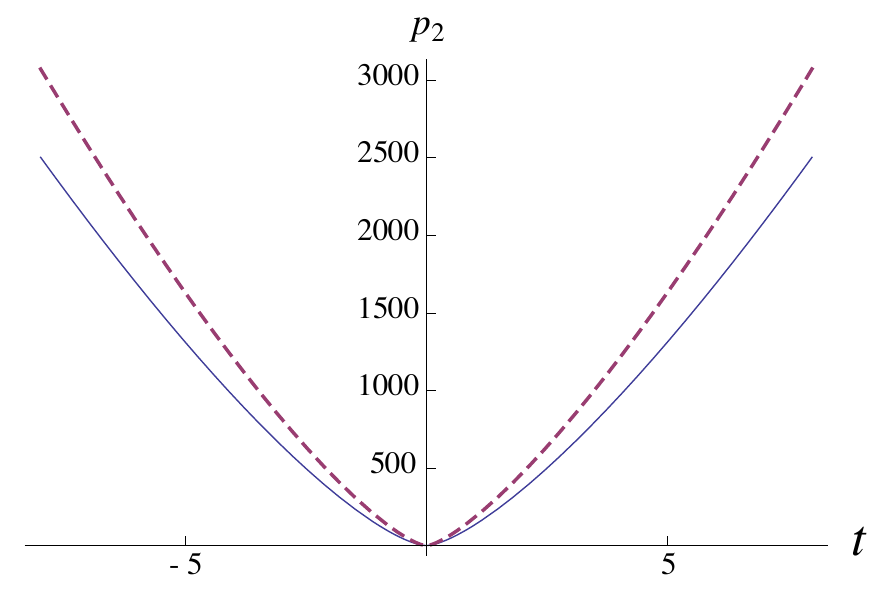}
\includegraphics*[height=3.6cm]{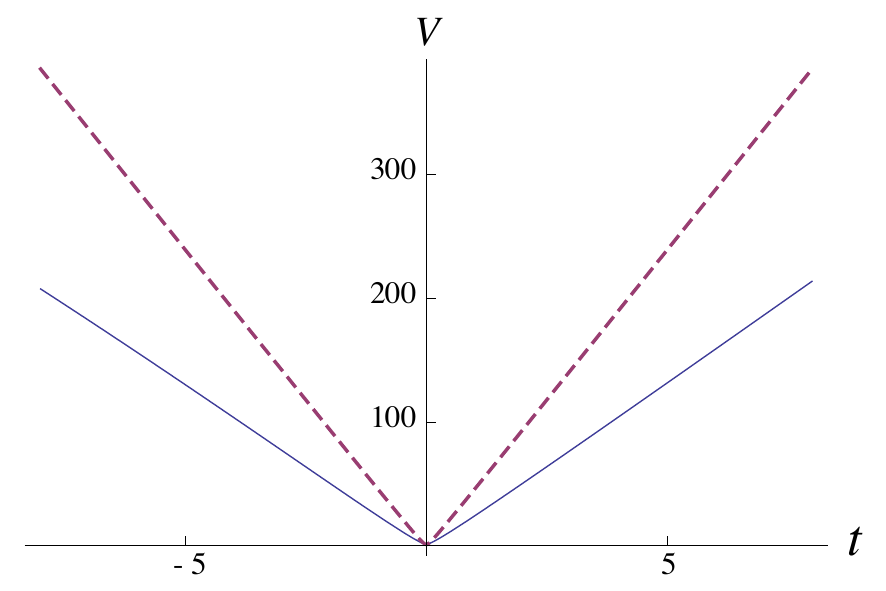}
\caption{ Left, middle and right panels  show $p_1$, $p_2$ and $V$ respectively for the Universe dominated by the massless vector field. Dashed red and blue lines denotes new and old $\bar{\mu}_i$ scheme respectively. It is worth to note, that even if $b(t_{cr})=b_{max}$ Ashtekar variables $p_1$ and $p_2$ have their global minima at $t=t_{cr}=0$.}
\label{fig:V1V2}
\end{figure}

\begin{figure}[h]
\includegraphics*[height=10.5cm]{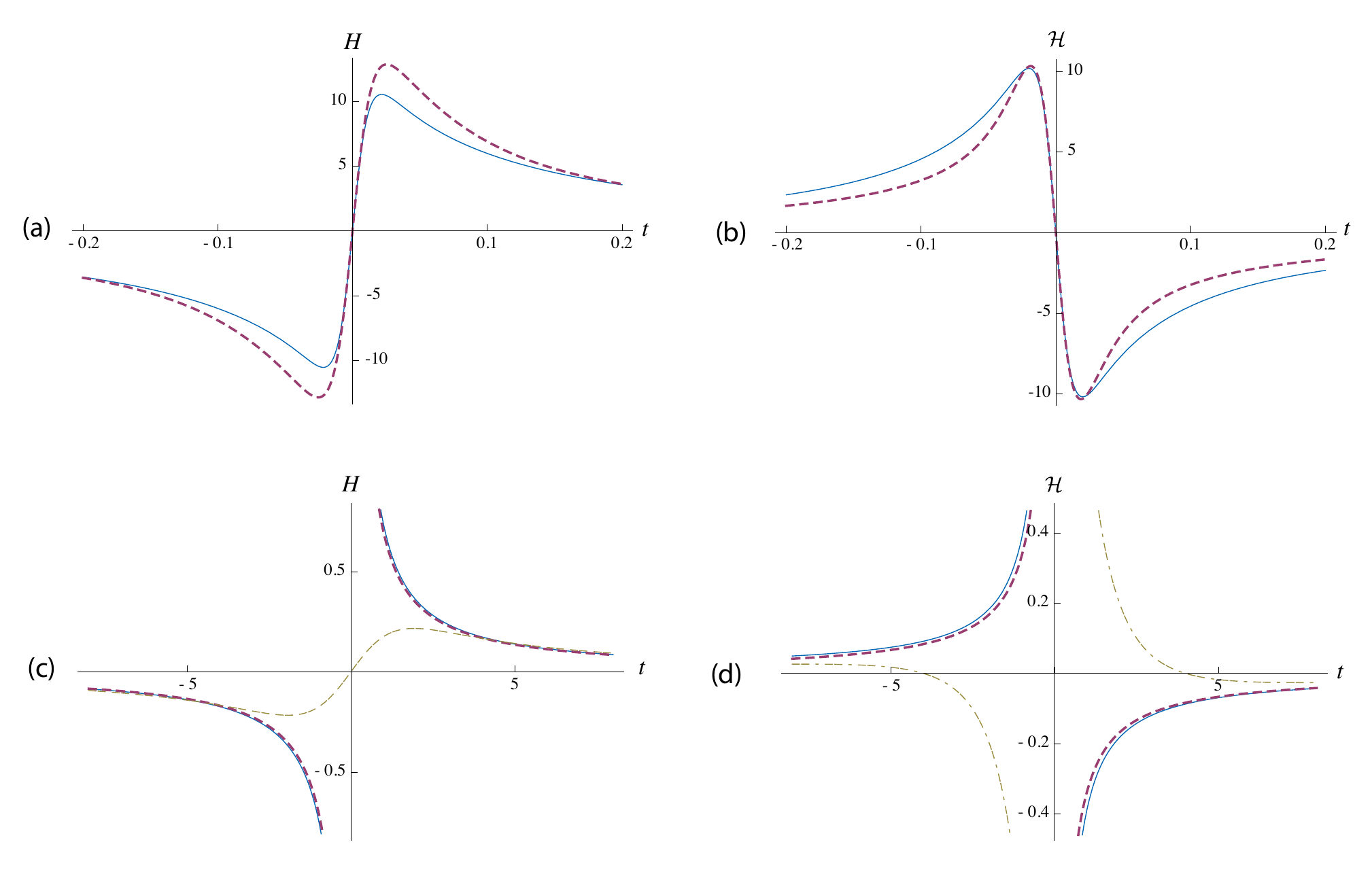}
\caption{ All panels show $H$ and $\mathcal{H}$ for the LQC with a massless vector field domination for the old (blue line) and the new (red dashed line) $\bar{\mu}_i$ scheme. For $t>0$ one founds $\mathcal{H}<0$ which means, that $b_{cr}$ is the global maximum of $b(t)$. Panels $(c)$ and $(d)$ show the LQC solution (blue and red lines) in a wider time range together with the GR limit (dashed yellow lines). In the panels (c) and (d) the moment of the Big Bounce is not shown for the LQC curves due to assumed scale of the plot.}
\label{fig:HaHbAbezmas}
\end{figure}

Referring to the LQC with a massless scalar field domination, let us consider a massless vector field in the LQC frame. Since the eq. (\ref{eq:rhooda}) is still valid one obtains $\rho=p\propto p_2^{-2}$. Numerical solutions for the scale factors are shown in the fig. \ref{fig:abklas-kwant}, Ashtekar parameters in fig. \ref{fig:V1V2}, Hubble parameters during the Big Bounce in fig. \ref{fig:HaHbAbezmas} and the low energy limit for both Hubble parameters \ref{fig:HaHbAbezmas}. The main difference with the classical scenario is the behaviour of $b(t)$ near the Planck scale. The contraction phase starts from $t=t_{cr}$, thus one obtains Kasner-like solution for $t\gg t_{cr}$. 
\\*

\subsection{Massive vector field domination}\label{sec:AmasaLQC}

The massless vector field domination leads to huge anisotropies and to contraction along the $z$ axis. To consider a more realistic scenario let us assume the massive vector field domination. We have assumed kinetic initial conditions for the vector field evolution, so for $\rho\sim\rho_{cr}$ one obtains solutions similar to the massless scenario. On the other hand the growing value of $A(t)$ stops the Kasner-like evolution, so $b_{cr}$ is only the local maximum of $b(t)$. Numerical calculations show, that for $m=10^{-5}m_{pl}$ one obtains $\mathcal{H}>0$ for $t>1000$ or $t>1500$ (old and new $\bar{\mu}_i$ scheme respectively).  The situation described in the subsection \ref{sec:masywnywektorKLAS} is a good low energy limit for the LQC evolution, so the decrease of anisotropy shown in the fig. \ref{fig:HiggsHaHb} remains valid in both schemes.
\\*

\section{LQC implications for realistic cosmological models}\label{sec:rozdzial3}

So far  we have studied the vector field domination in the early Universe. However, such scenarios can not produce inflation and lead to huge background anisotropies. Hence, let us turn to scenarios motivated by realistic inflationary models.

\subsection{Massive vector with a massive scalar}\label{sec:mvektorMskalarLQC}

Let us consider the Universe dominated by a vector field $A_\mu(t)$ with mass $m$ and a scalar field $\phi(t)$ with mass $M$. First of all we assume, that they are not coupled and that mass terms are time independent. Then one can introduce an $\alpha$ parameter defined by $\rho_\phi(t=t_{cr})=\alpha\rho_{cr}$, $\rho_A(t=t_{cr})=(1-\alpha)\rho_{cr}$. For numerical calculations we have assumed $\alpha=0.1$ (the vector field domination during the Big bounce) and $\alpha=0.5$ (vectors and scalars with the same contribution to $\rho_{cr}$). In both cases we assume $m=M=10^{-5}m_{pl}$. The evolution of the scale factor $b(t)$ is shown in the fig. \ref{fig:BB,masywne,b}. The ratio $\rho_A/\rho_\phi$ and evolution of $\mathcal{H}$ close to the Big Bounce are shown in fig. \ref{fig:BB,masywne,rhoA/rhophi,h}. The evolution of $b(t)$ for both values of $\alpha$ shows, that the presence of the scalar field reduces the period of Kasner-like evolution. On the other hand the vector field dominates for the initial period $t\sim O(10^3t_{pl})$, which leads to significant background anisotropies during that period.
\\*

\subsection{Vector mass from the Higgs mechanism}

Let us consider the vector field with the time dependent mass term produced by the Higgs mechanism together with a constant mass term. This strong coupling is motivated by the desired evolution of the vector field in the inflationary regime (see eq. (\ref{eq:ruchuU}) and discussion around it). The action of matter fields looks as follows:
\begin{equation}
S=\int d^4x\sqrt{-g}\left(\frac{1}{2}R-\frac{1}{4}F_{\mu\nu}F^{\mu\nu}+\frac{1}{2}\partial_\mu\phi\partial^\mu\phi+\frac{1}{2}(g(\phi)+m^2)A^\mu A_\mu-V(\phi)\right) \ . \label{eq:dzialanieAphi}
\end{equation}
It is easy to show, that the scalar field domination and $g(\phi)=-\frac{2}{3}V(\phi)$ give the slow-roll solution for $U(t)$ during inflation and the flat power spectrum of the initial inhomogeneities \cite{Dimopoulos:2006ms}. This makes it a useful choice, since often we would like the vector field to survive inflation and dominate the Universe after reheating. From the action (\ref{eq:dzialanieAphi}) one obtains the following equations of motion\footnote{Equations of motion can be also obtained from the scalar constraint (\ref{eq:Hamiltonian}) with $m^2\rightarrow m^2+g(\phi)$}
\begin{equation}
\ddot{\phi}+(2H+\mathcal{H})\dot{\phi}+V'\left(1-\frac{1}{3}\frac{A^2}{b^2}\right)=0 \ ,\qquad \ddot{A}+(2H-\mathcal{H})\dot{A}+(m^2-\frac{2}{3}V)A=0 \ . \label{eq:ruchuAphi}
\end{equation}
At the level of Einstein equations the strong coupling term contributes to $p_A$, so now $p_\phi=\frac{1}{2}\dot{\phi}^2-V$ and $p_A=\frac{1}{2b^2}(\dot{A}^2-(m^2-2V/3)A^2)$. Let us focuse on the LQC corrections to the evolution of this model. Since both, vector and scalar fields, are effectively massless close to the Big Bounce, then the strong coupling shall not have a significant influence on their evolution in that energy range. The evolution of scale factors is shown in the fig. \ref{fig:HiggsHaHb} The main deference appears for later times, when the extra mass term saves $A(t)$ from $\phi$ domination. Coupling to the scalar field has been introduced to cancel time dependent part of the effective mass for the vector in (\ref{eq:ruchuU}) and the result is that the amplitude of the vector field does not decreases faster than that of the scalar field. This can lead to the strong backreaction of the vector field on the evolution of space-time, which would give large background anisotropies. To avoid that one needs $U_\star <10^{-3}$, where $U_\star$ is the value of the physical vector field at the moment of horizon crossing of its perturbations. Larger $U_\star$ would produce measurable vector perturbations in the CMB. 

The other way to produce $m^2\simeq -2H^2$ is to consider a vector field with a non-minimal coupling to gravity. Unfortunately this model faces many problems in the LQC frame and so far we do not know how to to obtain effective equations of motion in this case.

\section{Conclusion}\label{sec:concl}

In the initial section of this paper the evolution of an anisotropic Universe dominated by vector field in GR frame has been studied. We have obtained the analytical solution for the massless vector field domination. In particular, it has been found, that there appears the quasi-Big Bounce solution, which is characterized by a finite value of the maximal energy density ($\rho(t_I)=\rho_I <\infty$), lack of initial curvature singularity ($R=0$ and $R^{\mu\nu}R_{\mu\nu}, R^{\mu\nu\alpha\beta}R_{\mu\nu\alpha\beta}<\infty$) and a non-zero initial value of the scale factor $a$ ($a(t_I)=a_I\neq 0$) or $H(t_I)=0$. On the other hand, along the direction set by the vector field one obtains characteristic features of the Big Bang for $t\rightarrow t_I$. The maximal energy density $\rho_I$ is a constant of integration. The $b(t)$ grows for $t\in (t_I,t_I+8/(3\sqrt{\rho_I}))$, and after that period one obtains Kasner-like solution. For the massive vector field domination one finds the isotropic Universe at late times.
\\*

In the section \ref{sec:rozdzial2} the LQC holonomy corrections to the simplified Bianchi I with a combination of vector and scalar fields have been considered. The effective equations of motion for Ashtekar variables, see sec. \ref{sec:ruchuLQC}-\ref{sec:nowy}, have been found for both, old and new $\bar{\mu}_i$ schemes. We have performed numerical simulations of the evolution of matter fields and scale factors for following scenarios: (i) massless vector field domination \ref{sec:AbezmasLQC}: we have obtained the Big Bounce at $t=t_{cr}$ and Kasner-like solution with decreasing value of $b(t)$. Thus $b(t_{cr})$ is the global maximum of $b(t)$. (ii) Massive vector field domination, see \ref{sec:AmasaLQC}: one obtains Kasner-like solution for the first $t\sim O(10^3 t_{pl})$, later on the Universe expands in all directions. 
\\*

In the final part of the paper we have discussed scenarios inspired by realistic inflationary models.  Domination of a combination of a vector and a scalar field, see \ref{sec:mvektorMskalarLQC}, produces the transition between contraction and expansion along the $z$ axis. Regardless of the initial contribution to the energy density, the vector field dominates the Universe, but one obtains the scalar field domination long before the vector oscillation phase. The relatively strong coupling between the massive vector and the massive scalar gives $\dot{b}>0$ for $t\sim O(0.1t_{pl})$. It is interesting to note, that non-trivial features on the evolution of the Universe appear in this case for energies much lower than the Planck scale. Also, it is important to note, that even during the $\dot{b}<0$ phase our model is consistent with the LQC, since both Ashtekar parameters reach their global minima at $t_{cr}$. One shall also note, that both forms of $\bar{\mu}_i$ functions gives similar results even at high energies, but this similarity can be at most qualitative, because of the scaling problem in the old $\bar{\mu}$ scheme. Future analysis shall be focused on the birth of quantum fluctuations in the anisotropic Universe with the LQC holonomy correction and on the non-minimal coupling of a vector field to gravity. Further research is needed to find out, how generic the results are.

\newpage
\begin{center}
{\bf Acknowledgements}
\end{center}
Authors thank Jerzy Lewandowski and Wojciech Kami{\'n}ski for useful discussions. This work was partially supported by Polish Ministry for Science and Education under grant N N202 091839.

\begin{figure}[h]
\centering
\includegraphics*[height=5cm]{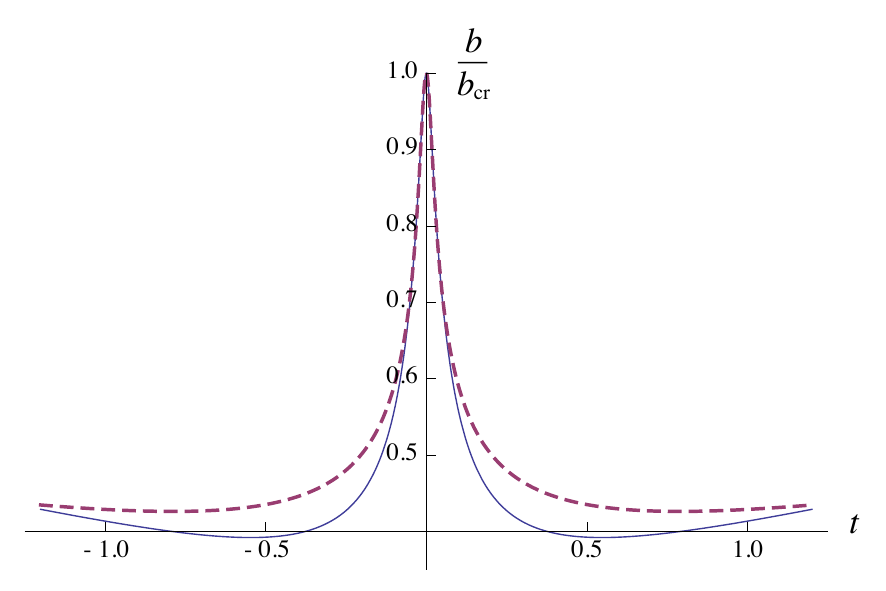}
\hspace{0.5cm}
\includegraphics*[height=5cm]{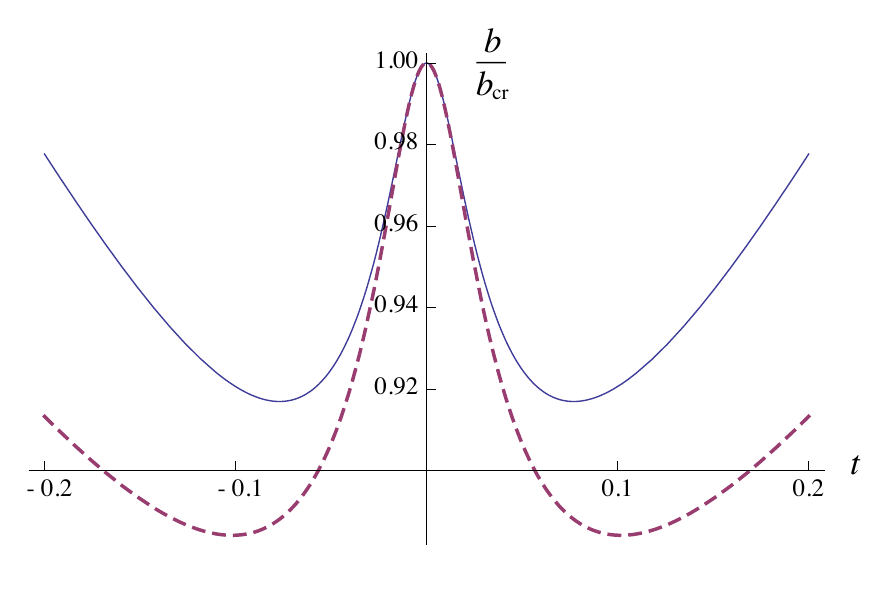}
\caption{ Left and right panels show $b(t)$ for LQC with the massive scalar and vector fields for $\alpha=0.5$ and $\alpha=0.1$respectively. Blue and dashed red lines denotes old and new $\bar{\mu}_i$ scheme. Initially $b(t)$ decrees, but after a while the scalar field starts to expand Universe in that direction. The $a(t)$ grows continuously from $a=a_{cr}$ like in the vector field domination scenario. One obtains very similar results for the vector mass coming from the Higgs mechanism.}
\label{fig:BB,masywne,b}
\end{figure}

\begin{figure}[h]
\centering
\includegraphics*[height=5cm]{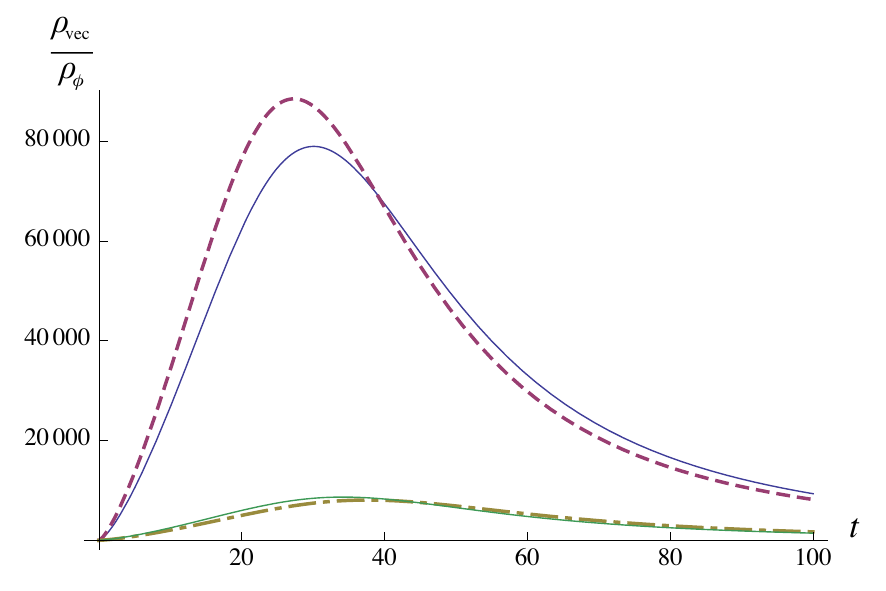}
\hspace{0.5cm}
\includegraphics*[height=5cm]{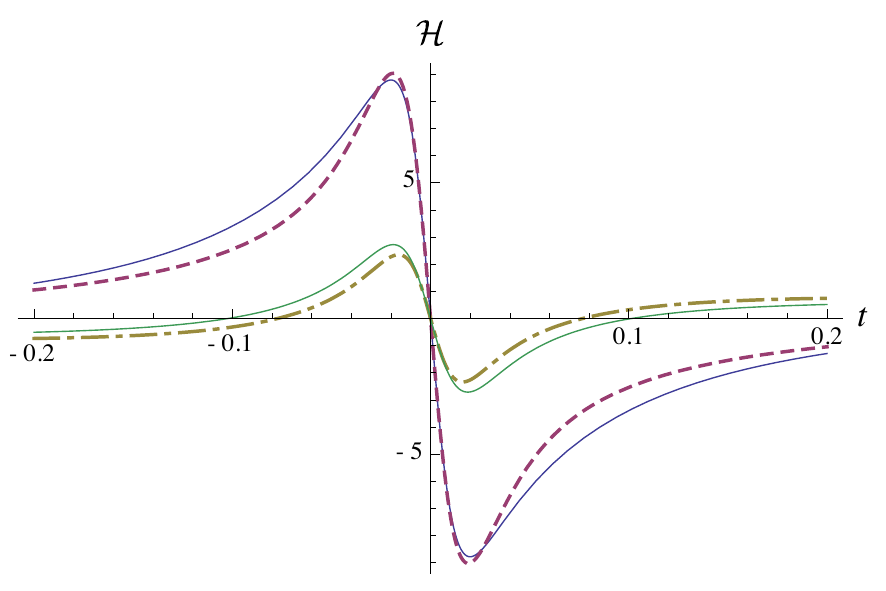}
\caption{ Left panel shows $\rho_A/\rho_\phi$ for LQC with the massive scalar and vector fields for $\alpha=0.1$ (blue and red lines) and $\alpha=0.5$ (yellow and green lines). The old and the new $\bar{\mu}_i$ scheme are denoted as continuous and dashed lines respectively. The vector field initially dominates the Universe, but after $10^3$ Planck times for $\alpha=0.5$ and after $4\times 10^3$ Planck times for $\alpha=0.1$ one obtains scalar field domination. The right panel shows the evolution of $\mathcal{H}$. For $\alpha=0.1$ one obtains $\mathcal{H}>0$ when $t>0.5 t_{pl}$}
\label{fig:BB,masywne,rhoA/rhophi,h}
\end{figure}

\end{document}